\documentclass[12pt]{article}
\usepackage{graphics,epsfig}
\newcommand{\beq}{\begin{equation}}
\newcommand{\eeq}{\end{equation}}
\newcommand{\bea}{\begin{eqnarray*}}
\newcommand{\eea}{\end{eqnarray*}}
\newcommand{\beaq}{\begin{eqnarray}}
\newcommand{\eeaq}{\end{eqnarray}}
\begin{document}

%\begin{flushright}APCTP-1999012\\KIAS-P99055\end{flushright}
\centerline{\Large\bf Exact One-Point Function of $N=1$}
\vskip .5cm
\centerline{\Large\bf super-Liouville Theory with Boundary}
\vskip 2cm
\centerline{\large Changrim Ahn$^{\dagger}$
\footnote{ahn@dante.ewha.ac.kr}, Chaiho Rim$^{\ddagger}$
\footnote{rim@phy.chonbuk.ac.kr}, 
and M. Stanishkov$^{\dagger}$
\footnote{stanishkov@dante.ewha.ac.kr; 
On leave of absence from Institute of Nuclear Research and Nuclear
Energy, Sofia, Bulgaria}
}
\vskip 1cm
\centerline{\it$^{\dagger}$Department of Physics, Ewha Womans University} 
\centerline{\it Seoul 120-750, Korea} \vskip .5cm
\centerline{\it$^{\ddagger}$ Department of Physics, Chonbuk National
University} \centerline{\it Chonju 561-756, Korea} \vskip 1cm
\centerline{\small PACS: 11.25.Hf, 11.55.Ds} \vskip 2cm
\centerline{\bf Abstract} 
In this paper, exact one-point functions of $N=1$ super-Liouville field 
theory in two-dimensional space-time with appropriate boundary conditions
are presented. Exact results are derived both for the theory defined on a 
pseudosphere with discrete (NS) boundary conditions and for the theory
with explicit boundary actions which preserves super conformal symmetries.
We provide various consistency checks. 
We also show that these one-point functions can be related to
a generalized Cardy conditions along with corresponding modular $S$-matrices.
Using this result, we conjecture the dependence of the boundary two-point 
functions of the (NS) boundary operators on the boundary parameter.

\newpage
\section{Introduction}

Liouville field theory (LFT) has been studied actively
for its relevance with non-critical string theories and two-dimensional
quantum gravity.
This theory is also interesting on its own as an example of irrational 
conformal field theory (CFT).
Most CFT formalism developed for rational CFTs do not apply to
this class of theories mainly because they have continuously infinite
number of primary fields.
Various methods have been proposed to derive structure
constants and reflection amplitudes, which are basic building
blocks to complete the conformal bootstrap \cite{Gervais,Teschner,ZamZam}.

Another example of the irrational CFT is the $N=1$ supersymmetric LFT (SLFT).
This model has some motivations.
It is applicable to the superstring theories and the 
two-dimensional supergravity with fermionic matter fields.
One can also understand the role of the extended conformal symmetry in the 
irrational CFTs by studing this model.
The methods introduced for the bulk LFT have been applied 
successfully to the bulk SLFT although the latter becomes algebraically 
more complicated \cite{RasSta,Poghossian}.

It is interesting to extend these formalisms to the CFTs 
defined in the two-dimensional
space-time geometry with a boundary condition (BC) which preserves
the conformal symmetry.
Cardy showed that the conformally invariant BCs 
can be associated with the primary fields in terms of 
modular $S$-matrix elements for the case of rational CFTs \cite{Cardy}.
It has been an issue whether the Cardy formalism can be extended to
the irrational CFTs.
Another motiviation is to understand open string theories 
in various nontrivial background space-time geometries \cite{string}. 

Recently it has been shown that functional relation method 
developed in \cite{Teschner} can be used in the boundary LFT 
with conformal boundary action
to derive one-point function of a bulk operator and correlation
functions of two boundary operators for a given conformal BC \cite{FZZ}. 
Here the conformal BC is denoted by a continuous parameter
related to the coupling constant in the boundary action. 
Another development made in \cite{ZamZam2} is to generalize
this method to the boundary LFT defined in the classical Lobachevskiy 
plane, namely the pseudosphere.
These works showed that the one-point functions of primary fields
can be related to the inner products of the conformal BCs and 
different Ishibashi states in the same way as the rational CFTs \cite{CarLew}.

It has been noticed that generalizing the Cardy formalism to the 
supersymmetric CFTs is nontrivial even for rational theories 
mainly because the Ramond (R) sector transforms not to Neveu-Schwarz (NS) 
sector but so-called (${\tilde{\rm NS}}$) sector 
under the modular transformation \cite{Apikyan,Nepomechie}.
In this paper we show that it is possible to derive exact one-point functions
from the functional relations and relate these to the Cardy formalism
for the BCs introduced in the pseudosphere and ordinary half plane
with conformally invariant boundary action.
This shows that the Cardy formalism can be generalized to the irrational
super-CFTs.
We also show that the boundary two-point functions of the 
(NS) boundary operators satisfy the same relation as those of the 
LFT.
This paper is organized as follows. In sect.2 we compute the one-point
functions of the SLFT primary fields on the pseudosphere. 
In sect.3 we propose the boundary action which preserves the superconformal
symmetry and compute the corresponding one-point functions.
These results are used in sect.4 to be related to the conformal BCs
and are shown to be consistent with the Cardy formalism.
We conclude the paper with some comments in sect.5.

%%%%%%%%%%%%%%%%%%%%%%%%%%%%%%%%%%%%%%%%%%%%%%%%%%%%%%%%%%%
\section{One-Point Functions of SLFT on a Pseudosphere}

The $N=1$ SLFT describes a supermultiplet consisting of a bosonic
field and its fermionic partner interacting with exponential potential. 
In terms of the component fields, the Lagrangian can be expressed by
\beq
{\cal L}_{\rm SL}={1\over{8\pi}}(\partial_{a}\phi)^2
-\frac{1}{2\pi}(\bar\psi\partial\bar\psi + \psi\bar\partial\psi)
+i\mu b^2\psi \bar\psi e^{b\phi} 
+{\pi\mu^2 b^2\over{2}}(:e^{b\phi}:)^2. 
\label{lagrangian}
\eeq
The first interaction term in Eq.(\ref{lagrangian}) containing 
$\exp(b\phi)$ has (holomorphic) conformal dimension $1/2$. 
This is correct since the Majorana fermion field has dimension $1/2$.
The SLFT is a superconformal field theory.
With the background charge $Q$
\beq
Q=b+{1\over{b}},
\eeq
the SLFT has the central charge
\beq
c_{SL}={3\over{2}}(1+2Q^2).
\eeq
The bulk (NS) and (R) primary fields of the SLFT are given by
\beq
N_\alpha(z,{\overline z})=e^{\alpha\phi},\qquad
R_\alpha(z,{\overline z})=\sigma^{(\epsilon)}(z,{\overline z})e^{\alpha\phi},
\eeq
where $\sigma^{(\pm)}$ is the `spin field' with dimension $1/16$
with even Fermi number for $(+)$ and odd for $(-)$
and satisfying an operator product expansion (OPE)
$\sigma^{(+)}\sigma^{(-)}=\psi$.
\footnote{We will suppress the superindex $(\epsilon)$ as much as we can
since the correlation functions of our interest do not depend on it
explicitly as we will show later.}
The conformal dimensions become, respectively,
\beq
\Delta^{N}_{\alpha}={1\over{2}}\alpha(Q-\alpha),\qquad
\Delta^{R}_{\alpha}=\frac{1}{16}+{1\over{2}}\alpha(Q-\alpha).
\eeq
From these, one can see that physical states can be denoted by
a real parameter $P$ defined by
\beq
\alpha={Q\over{2}}+iP.
\eeq

In this section we are interested in the SLFT on a pseudosphere.
This is a generalization of \cite{ZamZam2} where the LFT is studied in
the geometry of the infinite constant negative curvature surface, 
the so-called Lobachevskiy plane, i.e. the pseudosphere.
The equations of motion for the SLFT are given by
\beaq
\partial\bar\partial\phi &=& 4\pi^2\mu b^2\left( \mu
e^{b\phi}-i\bar\psi\psi\right)e^{b\phi}\\
\partial\bar\psi &=& -i\mu e^{b\phi}\psi,\qquad
\bar\partial\psi= i\mu e^{b\phi}\bar\psi.
\eeaq
We will assume that the fermion vanishes in the classical limit so that
the background metric is determined by the bosonic field satisfying
\beq
e^{\varphi(z)}={4R^2\over{(1-|z|^2)^2}},
\eeq
where $\varphi=2b\phi$ and $R^{-2}=4\pi^2\mu^2 b^3$.
The parameter $R$ is interpreted as the radius of the pseudosphere
in which the points at the circle $|z|=1$ are infinitely
far away from any internal point.
This circle can be interpreted as the ``boundary'' of the pseudosphere.
In the same way as the LFT, we can now use the Poincar\'e model of
the Lobachevskiy plane with complex coordinate $\xi$ in the 
upper half plane.

We want to compute exact one-point functions of the (NS) and (R)
bulk operators $N_\alpha$ and $R_\alpha$. 
Due to the superconformal invariance, these one-point functions are given by
\beaq
\langle N_\alpha(\xi)\rangle
&=& {U^N(\alpha)\over |\xi-\bar\xi|^{2\Delta^N_\alpha}}\\
\langle R_\alpha(\xi)\rangle
&=& {U^R(\alpha)\over |\xi-\bar\xi|^{2\Delta^R_\alpha}}.
\eeaq
We will simply refer to the coefficients $U^N(\alpha)$ and $U^R(\alpha)$ as 
bulk one-point functions.
To derive the functional relations satisfied by these one-point
functions, we should consider the bulk degenerate fields 
which are defined by some differential equations
with certain orders \cite{FQS}.

The degenerate fields in the (NS) sector are given by
\beq
N_{\alpha_{m,n}}=e^{\alpha_{m,n}\phi},\quad
\alpha_{m,n}={1\over{2b}}(1-m)+{b\over{2}}(1-n),
\quad{\rm with}\quad m-n={\rm even}
\label{nsdeg}
\eeq
and those in the (R) sector by
\beq
R_{\alpha_{m,n}}=\sigma^{(\epsilon)} e^{\alpha_{m,n}\phi},\quad
{\rm with}\quad m-n={\rm odd}.
\eeq
One of the essential features of these fields is that the operator
product expansion (OPE) of a degenerate field with any primary
field is given by a linear combination of only finite
number of primary fields and their decendents.
The simplest degenerate fields are $N_{-b}$ for the (NS) sector and
$R_{-b/2}$ for the (R) sector.
In particular, $R_{-b/2}$ satisfies 
\beq
\left[{1\over{b^2}}\partial+G_{-1}G_{0}\right]R_{-b/2}=0
\eeq
where $G$ is the fermionic part of the supercurrent.
Similar equation holds for the anti-holomorphic part. 

The OPE of $R_{-b/2}$ with a (NS) primary field is given by
\begin{equation}
N_{\alpha}R_{-b/2}=C^{(N)}_{+}(\alpha)\left[R_{\alpha-b/2}\right]
+C^{(N)}_{-}(\alpha)\left[R_{\alpha+b/2}\right],
\label{nsope}
\end{equation}
where $[\ldots]$ stands for entire family of conformal decendents
corresponding to a primary field.
The structure constants can be computed 
using Coulomb gas integrals.
One can set $C^{(N)}_{+}=1$ since no screening insertion is needed. 
The other structure constant $C^{(N)}_{-}$ needs just one 
insertion of the SLFT interaction and can be obtained by
\beaq
C^{(N)}_{-}(\alpha)&=&-i\mu b^2\int d^2\xi\langle N_{\alpha}(0)R_{-b/2}(1)
\psi(\xi)\bar\psi({\overline\xi}) 
e^{b\phi(\xi,{\overline\xi})}R_{Q-\alpha-b/2}(\infty)\rangle\nonumber\\
&=&{\mu b^2\over{2}} \int d^2{\xi} |\xi|^{-2\alpha b}|1-\xi|^{b^2-1}=
{\pi\mu b^2\gamma\left(\alpha b-{b^2\over{2}}
-{1\over{2}}\right)\over{2\gamma\left({1-b^2\over{2}}\right)\gamma(\alpha b)}}
\eeaq
with $\gamma(x)=\Gamma(x)/\Gamma(1-x)$.
Here, we used the following correlation functions of the spin fields
\beaq
\langle\sigma(\xi_1,{\overline \xi_1})
\sigma(\xi_2,{\overline \xi_2})\rangle&=&
{1\over{|\xi_{12}|^{1/4}}}\\
\langle\sigma(\xi_1,{\overline \xi_1})\sigma(\xi_2,{\overline \xi_2})
\psi(\xi_3){\overline\psi}({\overline \xi_3})\rangle&=&
{i/2\over{|\xi_{12}|^{-3/4}|\xi_{13}||\xi_{23}|}}.
\eeaq

Similarly, the OPE with the (R) primary field is
\begin{equation}
R_{\alpha}R_{-b/2}=C^{(R)}_{+}(\alpha)\left[N_{\alpha-b/2}\right]
+C^{(R)}_{-}(\alpha)\left[N_{\alpha+b/2}\right]
\label{rope}
\end{equation}
where $C^{(R)}_{+}=1$ as before and
$C^{(R)}_{-}$ is given by a screening integral
\beaq
C^{(R)}_{-}(\alpha)&=&-i\mu b^2\int d^2\xi\langle R_{\alpha}(0)R_{-b/2}(1)
\psi(\xi)\bar\psi({\overline\xi}) 
e^{b\phi(\xi,{\overline\xi})}N_{Q-\alpha-b/2}(\infty)\rangle\nonumber\\
&=&{\mu\over{2}} b^2\int d^2 \xi |\xi|^{-2\alpha b-1}|1-\xi|^{b^2-1}=
{\mu\over{2}} b^2{\pi\gamma\left(\alpha b-{b^2\over{2}}\right)\over{
\gamma\left({1-b^2\over{2}}\right)\gamma\left(\alpha b+{1\over{2}}\right)}}.
\label{cminus}
\eeaq

Now we consider the bulk two-point functions of 
the degenerate field $R_{-b/2}$ and a (NS) field $N_\alpha$,
\beq
G^{N}_{\alpha,-b/2}(\xi,\xi')=\langle N_{\alpha}(\xi')R_{-b/2}(\xi)\rangle.
\eeq
It is straightforward from Eq.(\ref{nsope}) to show that
the two-point function satisfy
\beq
\label{nsone}
G^{N}_{\alpha,-b/2}(\xi,\xi')
=C^{(N)}_{+}(\alpha)U^{R}\left(\alpha-{b\over{2}}\right)
{\cal G}_{+}(\xi,\xi')+C^{(N)}_{-}(\alpha)U^{R}\left(\alpha+{b\over{2}}\right)
{\cal G}_{-}(\xi,\xi') 
\eeq
where ${\cal G}_{\pm}(\xi,\xi')$ are expressed in terms of the special
conformal blocks
\beq
{\cal G}_{\pm}(\xi,\xi')={|\xi'-{\overline \xi'}|^{2\Delta^{N}_{\alpha}
-2\Delta^{R}_{-b/2}}\over{
|\xi-{\overline \xi'}|^{4\Delta^{N}_{\alpha}}}}{\cal F}_{\pm}(\eta)
\eeq
with 
\beq
\eta={(\xi-\xi')({\overline \xi}-{\overline \xi'})\over{(\xi-{\overline \xi'})
({\overline \xi}-\xi')}}.
\eeq
Here, the conformal blocks are given by the hypergeometric functions
\beaq
{\cal F}_{+}(\eta)&=&\eta^{{\alpha b\over{2}}}
(1-\eta)^{-{b^2\over{4}}+{3\over{8}}}
F\left(\alpha b-b^2,{1\over{2}}-{b^2\over{2}};\alpha b-{b^2\over{2}}+
{1\over{2}};\eta\right),\\
{\cal F}_{-}(\eta)&=&\eta^{{1\over{2}}+{b^2\over{2}}-{\alpha b\over{2}}}
(1-\eta)^{-{b^2\over{4}}+{3\over{8}}}
F\left({1\over{2}}-{b^2\over{2}},1-\alpha b;-\alpha b+{b^2\over{2}}+
{3\over{2}};\eta\right).
\eeaq
In the cross channel, an equivalent expression for the two-point
function can be obtained as follows:
\beq
G^{N}_{\alpha,-b/2}={|\xi'-\bar\xi'|^{2\Delta^N_\alpha-2\Delta^R_{-b/2}}
\over |\xi-\bar\xi'|^{4\Delta^N_\alpha}}
\left[B^{(N)}_{+}(\alpha){\widetilde{\cal F}_{+}}(\eta)
+B^{(N)}_{-}(\alpha){\widetilde{\cal F}_{-}}(\eta)\right]
\label{nscross} 
\eeq
with
\beaq
{\widetilde{\cal F}_{+}}(\eta)&=& \eta^{{\alpha b\over{2}}}
(1-\eta)^{-{b^2\over{4}}+{3\over{8}}}
F\left(\alpha b-b^2,{1\over{2}}-{b^2\over{2}};1-b^2;1-\eta\right),\\
{\widetilde{\cal F}_{-}}(\eta)&=&
\eta^{{\alpha b\over{2}}}
(1-\eta)^{{3b^2\over{4}}+{3\over{8}}}
F\left({1\over{2}}+{b^2\over{2}},\alpha b;1+b^2;1-\eta\right).
\eeaq
The boundary structure constants $B^{(N)}_{\pm}$ can be determined
from the monodromy relations
\beaq
{\cal F_+}({\eta}) &=& {\Gamma\left(\alpha b-{b^2\over{2}}+{1\over{2}}\right)
\Gamma(b^2)\over\Gamma\left({b^2\over{2}}+{1\over{2}}\right)
\Gamma(\alpha b)}{\widetilde{\cal F_+}}({\eta})
+{\Gamma\left(\alpha b-{b^2\over{2}}+{1\over{2}}\right)
\Gamma(-b^2)\over \Gamma(\alpha b-b^2) 
\Gamma\left({1\over{2}}-{b^2\over{2}}\right)}
{\widetilde{\cal F_-}}({\eta}),\\
{\cal F_-}({\eta}) &=& {\Gamma\left(-\alpha b+{b^2\over{2}}+{3\over{2}}\right)
\Gamma(b^2)\over\Gamma\left({b^2\over{2}}+{1\over{2}}\right)
\Gamma(1+b^2-\alpha b)}{\widetilde{\cal F_+}}({\eta})
+{\Gamma\left(-\alpha b+{b^2\over{2}}+{3\over{2}}\right)
\Gamma(-b^2)\over \Gamma(1-\alpha b)
\Gamma\left({1\over{2}}-{b^2\over{2}}\right)}{\widetilde{\cal F_-}}({\eta}).
\eeaq

The conformal block $\widetilde{\cal F_-}$ corresponds to the identity
boundary operator with dimension $0$ appearing in the boundary as
the bulk operator $R_{-b/2}$ approaches the boundary with $\eta\rightarrow 1$.
Another boundary operator $n_{-b}$ appearing as $R_{-b/2}$ approaches
the boundary generates the $\widetilde{\cal F_+}$ block.
As mentioned above, the geodesic distance to the boundary
on the pseudosphere is infinite.
Therefore, the two-point function in the LHS of (\ref{nsope}) can be
factorized into a product of two one-point functions and satisfies
\beq
B^{(N)}_{-}(\alpha)=U^N(\alpha)U^R(-b/2).
\eeq
Combining all these and using (\ref{nscross}), we obtain the
following nonlinear functional equation in the $\eta\rightarrow 1$ limit:
\beq
{\Gamma\left({1-b^2\over{2}}\right)U^{N}(\alpha)U^R\left(-{b\over{2}}\right)
\over{\Gamma(-b^2)\Gamma\left(\alpha b-{b^2\over{2}}+{1\over{2}}\right)}}
={U^{R}\left(\alpha-{b\over{2}}\right)\over{\Gamma\left(\alpha b-b^2\right)}}
+{\pi\mu b^2\ U^{R}\left(\alpha+{b\over{2}}\right)
\over{\gamma\left({1-b^2\over{2}}\right)\Gamma(\alpha b)
\left(\alpha b-{b^2\over{2}}-{1\over{2}}\right)}}.
\label{firsteq}
\eeq

Analysis of the other two-point function
\beq
G^{R}_{\alpha,-b/2}(\xi,\xi')=\langle R_{\alpha}(\xi)R_{-b/2}(\xi')\rangle
\label{rtwopt}
\eeq
goes along the same line and leads to the second functional equation
\beq
{\Gamma\left({1-b^2\over{2}}\right)U^{R}(\alpha)U^R\left(-{b\over{2}}\right)
\over \Gamma(-b^2)\Gamma\left(\alpha b-{b^2\over{2}}\right)}=
{U^{N}\left(\alpha-{b\over{2}}\right)\over
\Gamma\left(\alpha b-b^2-{1\over{2}}\right)}
+{\pi\mu b^2\ U^{N}\left(\alpha+{b\over{2}}\right)\over{
\gamma\left({1-b^2\over{2}}\right)\Gamma\left(\alpha b+{1\over{2}}\right)}}.
\label{secondeq}
\eeq
Here, we used the following (R)-sector conformal blocks \cite{MSS}
\bea
{\cal F}^{R}_{+}(\eta)&=&\eta^{{\alpha b\over{2}}+{1\over{8}}}
(1-\eta)^{-{b^2\over{4}}+{3\over{8}}}
\left(1+\sqrt{1-{1\over{\eta}}}\right)^{1/2}
F\left(\alpha b-b^2-{1\over{2}},{1-b^2\over{2}};\alpha b-{b^2\over{2}}
;\eta\right)\\
{\cal F}^{R}_{-}(\eta)&=&\eta^{{b^2\over{2}}-{\alpha b\over{2}}-{5\over{8}}}
(1-\eta)^{-{b^2\over{4}}-{1\over{8}}}
\left(1-\sqrt{1-{1\over{\eta}}}\right)^{1/2}
F\left({-1-b^2\over{2}},{1\over{2}}-\alpha b;-\alpha b+{b^2\over{2}}+
1;\eta\right).
\eea
It is worth mentioning on the superindices $(\epsilon)$ we have omitted
in the correlation function (\ref{rtwopt}).
When one (R) field has $\epsilon=+$ and the other $\epsilon=-$,
the two-point function becomes proportional to 
$\langle \sigma^{(+)}\sigma^{(-)}\rangle$ or $\langle\psi\rangle$.
And the fermion one-point function should vanish due to the Lorenz invariance.
If both (R) fields have the same index, either $\epsilon=+$ or $-$,
the previous derivation holds as it is and the same functional relations
are obtained.

The SLFT satisfies the duality $b\to 1/b$. 
This property requires considering another degenerate (R) operator
$R_{-1/2b}$ which generates two more functional equations
in addition to Eqs.(\ref{firsteq}) and (\ref{secondeq}).
These additional equations can be obtained by just replacing the 
coupling constant $b$ with $1/b$ and the paramter $\mu$ by
the ``dual'' $\widetilde\mu$ satisfying
\beq
\pi\widetilde\mu\gamma\left({Q\over 2b}\right)=
\left[\pi\mu\gamma\left({bQ\over 2}\right) \right]^{1/{b^2}}.
\label{dualmu}
\eeq
Therefore, the one-point functions $U^{N}(\alpha)$ and $U^{R}(\alpha)$
should satisfy four nonlinear functional equations.

We have found the solutions to these overdetermined nonlinear equations
as follows:
\beaq
U^{N}_{mn}(\alpha)&=&{\sin\left({\pi Q\over{2b}}\right)
\sin\left({\pi bQ\over{2}}\right)
\sin\left[m\pi\left({Q\over{2b}}-{\alpha\over{b}}\right)\right]
\sin\left[n\pi\left({bQ\over{2}}-b\alpha\right)\right]
\over{
\sin\left({m\pi Q\over{2b}}\right)
\sin\left({n\pi bQ\over{2}}\right)
\sin\left[\pi\left({Q\over{2b}}-{\alpha\over{b}}\right)\right]
\sin\left[\pi\left({bQ\over{2}}-b\alpha\right)\right]}} U^{N}_{11}(\alpha)
\label{unmn}\\
U^{R}_{mn}(\alpha)&=&{\sin\left({\pi Q\over{2b}}\right)
\sin\left({\pi bQ\over{2}}\right)
\sin\left[m\pi\left({Q\over{2b}}-{\alpha\over{b}}+{1\over{2}}\right)\right]
\sin\left[n\pi\left({bQ\over{2}}-b\alpha+{1\over{2}}\right)\right]
\over{ \sin\left({m\pi Q\over{2b}}\right)\sin\left({n\pi bQ\over{2}}\right)
\cos\left[\pi\left({Q\over{2b}}-{\alpha\over{b}}\right)\right]
\cos\left[\pi\left({bQ\over{2}}-b\alpha\right)\right] }} U^{R}_{11}(\alpha)
\quad
\label{urmn}
\eeaq
where the `basic' solutions are given by
\beaq
U^{N}_{11}(\alpha)&=&
\left[\pi\mu\gamma\left({bQ\over{2}}\right)\right]^{-\alpha/b}
{\Gamma\left({bQ\over{2}}\right)\Gamma\left({Q\over{2b}}\right){Q\over{2}}
\over{ \Gamma\left(-\alpha b+{bQ\over{2}}\right)
\Gamma\left(-{\alpha\over{b}}+{Q\over{2b}}\right)
\left({Q\over{2}}-\alpha\right) }}
\label{un11}\\
U^{R}_{11}(\alpha)&=& 
\left[\pi\mu\gamma\left({bQ\over{2}}\right)\right]^{-\alpha/b}
{\Gamma\left({bQ\over{2}}\right)\Gamma\left({Q\over{2b}}\right){Q\over{2}}
\over{ \Gamma\left(-\alpha b+{bQ\over{2}}+{1\over{2}}\right)
\Gamma\left(-{\alpha\over{b}}+{Q\over{2b}}+{1\over{2}}\right)}}.
\label{ur11}
\eeaq
This is our main result in this section. 
There are infinite number of possible solutions which are parametrized
by two integers $(m,n)$.
For these to be solutions, we find that 
the two integers should satisfy $m-n=$ even.
The basic solutions, Eqs.(\ref{un11}) and (\ref{ur11}),
can be interpreted as the one-point functions of the bulk
operators $N_{\alpha}$ and $R_{\alpha}$ corresponding to the vacuum BC,
the BC corresponding to the bulk vacuum operator $N_{0}$.
Then, the  general solutions, Eqs.(\ref{unmn}) and (\ref{urmn}), 
can be identified with the one-point functions with the conformal
BC $(m,n)$ classified by Cardy \cite{Cardy}.
We will discuss more about this issue in sect.4. 
Since $m-n=$ even, the one-point functions we obtained correspond to the 
(NS)-type BCs only.
This seems consistent with the fact that only the (NS) boundary
operators arise when the (NS) or (R) bulk degenerate operators approach 
the boundary corresponding to the vacuum BC.

We also note that Eqs.(\ref{unmn}) and (\ref{urmn}) satisfy 
so-called bulk ``reflection relations'':
\beq
\label{bulkrefrel}
U^{N}_{m,n}(\alpha)=D^{(N)}(\alpha)U^{N}_{m,n}(Q-\alpha),\quad
U^{R}_{m,n}(\alpha)=D^{(R)}(\alpha)U^{R}_{m,n}(Q-\alpha)
\eeq
where $D^{(N)}(\alpha)$ and $D^{(R)}(\alpha)$ are the (NS) and 
the (R) reflection amplitudes derived in \cite{RasSta,Poghossian}.
One can also notice that new functions defined by
\beq
{\widetilde U}^{N}_{m,n}=U^{N}_{m,n},\qquad
{\widetilde U}^{R}_{m,n}=-U^{R}_{m,n}
\label{utilde}
\eeq
become again solutions to the above nonlinear equations.
We will discuss the meaning of the solutions from the viewpoint of the 
generalized Cardy formalism later.

Finally, we consider a special case where $\alpha=0$. 
From Eq.(\ref{unmn}), one can notice that one-point function of 
the identity operator is normalized in such a way that $U^{N}(0)=1$.
Then, one-point function of the spin field with $(m,n)$ BC is given by
Eq.(\ref{urmn})
\beq
\langle\sigma^{(\pm)}\rangle_{(m,n)}=U^{R}_{mn}(0).
\eeq

\section{Bulk One-Point Function with Boundary}

In this section, we define the SLFT on half plane where superconformally
invariant boundary action is imposed.  
We choose the following boundary action at $y=0$
\begin{equation}
{\cal L}_{B}={\mu_B\over{2}}
e^{b\phi/2}a(\psi-i\gamma{\overline\psi})(x)
\label{baction}
\end{equation}
with $\gamma=\pm 1$ and the fermionic zero-mode $a$ satisfying \cite{GZ}
\beq
\sigma^{(\pm)}=a\sigma^{(\mp)}\qquad{\rm and}\qquad
a^2=1.
\eeq
This action includes additional boundary parameter $\mu_B$ which
generates continuous family of the BCs.
The boundary equations of motion are given by
\beaq
{1\over{2\pi}}\partial_{y}\phi&=&-{1\over{2}}b\mu_B a
(\psi-i\gamma{\overline\psi})e^{b\phi/2}\\
{i\over{2\pi}}\psi&=&\mu_B e^{b\phi/2}\ a,\qquad
{i\over{2\pi}}{\overline \psi}=i\gamma\mu_B e^{b\phi/2}\ a
\eeaq
which lead to
\beq
(\psi+i\gamma{\overline\psi})=0.
\label{cardyeq}
\eeq 
Notice that Eq.(\ref{cardyeq}) is the well-known fermion 
BC imposed by Cardy.
Plugging these constraints back into the action, one can
simplify the boundary action 
\beq
{\cal L}_{B}=\mu_B e^{b\phi/2}a\psi.
\eeq
This term is different from that preserving boundary integrability 
such as considered in \cite{Prata}.
Main difference is that our action is preserving not only integrability 
but also superconformal symmetry.
Indeed, the boundary action is nothing but the screening operator
which guarantees the symmetry.

One can see that physical quantities should contain only even powers of 
$\mu_B$ because of the fermionic zero-mode.
While the bulk properties of the boundary SLFT should be identical,
we should define the boundary operators.
As in the bulk, there are two sectors, the (NS) and (R) boundary operators
\beq
n_{\beta}=e^{\beta\phi/2}(x),\qquad
r_{\beta}=\sigma^{(\epsilon)} e^{\beta\phi/2}(x).\qquad
\eeq

The procedure to derive the functional equations satisfied by
the bulk one-point functions are identical to that in sect.2.
Major difference arises when the bulk degenerate operator $R_{-b/2}$ 
approaches the boundary as $z\to{\overline z}$.
The LHS of Eq.(\ref{nsone}) can be evaluated by the boundary OPE which
generates the boundary operator $n_0$ and $n_{-b}$.
We choose the identity operator $n_0$, or the boundary vacuum state,
since we are interested in the bulk one-point function. 
The fusion of the degenerate field $R_{-b/2}$ can be computed by 
a first order perturbation from the boundary action:
\beaq
{\cal R}^{(\epsilon)}(-b/2,Q)&=&-\mu_B
\int dx\langle R^{(\epsilon)}_{-b/2}\left({i\over{2}}\right) a\psi(x)
e^{b\phi_B /2}(x) e^{Q\phi_B /2}(\infty)\rangle
\nonumber\\
&=&\mu_B\int dx |x-i/2|^{b^2-1}=
2\pi\mu_B{\Gamma(-b^2)\over{\Gamma\left({1-b^2\over{2}}\right)^2}}.
\eeaq
Here, we used the formula
\beq
\langle \sigma(\xi,{\overline\xi})a\psi(x)\rangle={-1\over{|x-\xi|^{1/2}
|\xi-{\overline\xi}|^{3/8}}}.
\eeq
Again, the dependence on the superindex $\epsilon$ disappears so that 
we can suppress it.
With the vacuum state on the boundary, the two-point function becomes 
the bulk one-point function of the operator $N_{\alpha}$. 
Equating this with the RHS of Eq.(\ref{nsone}) gives the 
functional equation
\beq
{2\pi\mu_{B}\over{\Gamma\left({1-b^2\over{2}}\right)}}U^{N}(\alpha)
={\Gamma\left(\alpha b-{b^2\over{2}}+{1\over{2}}\right)\over{
\Gamma\left(\alpha b-b^2\right)}}U^{R}\left(\alpha-{b\over{2}}\right)
+{\pi\mu b^2\Gamma\left(\alpha b-{b^2\over{2}}-{1\over{2}}\right)\over{
\gamma\left({1-b^2\over{2}}\right)\Gamma(\alpha b)}}
U^{R}\left(\alpha+{b\over{2}}\right).
\label{eqns}
\eeq
Similar consideration for the $G^{R}_{\alpha,-b/2}$ leads to
\beq
{2\pi\mu_{B}\over{\Gamma\left({1-b^2\over{2}}\right)}}U^{R}(\alpha)
={\Gamma\left(\alpha b-{b^2\over{2}}\right)\over{
\Gamma\left(\alpha b-b^2-{1\over{2}}\right)}}
U^{N}\left(\alpha-{b\over{2}}\right)
+ {\pi\mu b^2 \Gamma\left(\alpha b-{b^2\over{2}}\right)\over{
\gamma\left({1-b^2\over{2}}\right)\Gamma\left(\alpha b+{1\over{2}}\right)}}
U^{N}\left(\alpha+{b\over{2}}\right).
\label{eqr}
\eeq
As before, one should consider the dual equations coming from the dual 
degenerate operator $R_{-1/2b}$.

The solutions of Eqs.(\ref{eqns}) and (\ref{eqr}) can be found as
\beaq
\label{UN}
&U^{N}(\alpha)={\cal N}b\left[\pi\mu\gamma\left({bQ\over{2}}\right)
\right]^{{Q-2\alpha\over{2b}}}
\Gamma\left(\left(\alpha-{Q\over{2}}\right)b\right)
\Gamma\left(1+\left(\alpha-{Q\over{2}}\right){1\over{b}}\right)
\cosh\left[\left(\alpha-{Q\over{2}}\right)\pi s\right]\\
\label{UR} 
&U^{R}(\alpha)={\cal N}\left[\pi\mu\gamma\left({bQ\over{2}}\right)
\right]^{{Q-2\alpha\over{2b}}}
\Gamma\left(\left(\alpha-{b\over{2}}\right)b\right)
\Gamma\left(\left(\alpha-{1\over{2b}}\right){1\over{b}}\right)
\cosh\left[\left(\alpha-{Q\over{2}}\right)\pi s\right],
\eeaq
where the normalization factor ${\cal N}$ is given by
\beq
{\cal N}=\left[\pi\mu\gamma\left({bQ\over{2}}\right)
\right]^{-Q/2b}
\left[b\Gamma(-Qb/2)\Gamma(1-Q/2b)\cosh(Q\pi s/2)\right]^{-1}
\eeq
so that $U^{N}(0)=1$.
Here, the boundary parameter $s$ is related to $\mu_B$ by
\beq
{\mu_B^2\over{\mu b^2}}\sin\left({\pi bQ\over{2}}\right)=\cosh^2\left(
{\pi bs\over{2}}\right).
\eeq

It is possible to find another conformal BC by changing $a\to -a$.
This introduces extra `--' sign in the LHS of the functional equations,
Eqs.(\ref{eqns}) and (\ref{eqr}), so that the solutions are found to be
\beq
{\widetilde U}^{(N)}=U^{(N)},\qquad {\widetilde U}^{(R)}=-U^{(R)}.
\label{utildec}
\eeq
The solutions Eqs.(\ref{UN}) and (\ref{UR}) are our main result in this
section.  Notice that these are self-dual if the parameter $s$ is 
invariant and $\mu\to{\widetilde\mu}$ as Eq.(\ref{dualmu}).
The continuous parameter $s$ coming from $\mu_B$ generates a continuous
family of conformally invariant BCs. We will discuss how these BCs
can be consistent with the generalized Cardy formalism in the next section.
One can also check that these satisfy the bulk reflection relations 
Eq.(\ref{bulkrefrel}).

The one-point function can be checked by a perturbative analysis.
Defining the third and fourth terms in Eq.(\ref{lagrangian}) as 
$V_b$ and the boundary action in Eq.(\ref{baction}) as $B_b$,
one can express an one-point function as an infinite sum of these
perturbation terms,
\beq
\langle O_{\alpha} (\xi, \bar \xi)\rangle
=\sum_{p,q} \frac 1 {p!q!}
\langle O_{\alpha}(\xi, \bar\xi) \, V_b^p \, B_b^q \rangle_0
\eeq
where the evaluation is made with $\mu=\mu_B=0$.
It is well known that the perturbative results are non-vanishing
only at the on-shell condition $\alpha = Q/2 - (p+ q/2)b$
and correspond to the residue of the one-point function as follows:
\beaq
{\rm residue }  \,\, U^N (\alpha) |_{\alpha =  Q/2 - (p+q/2)b}
&= &{\cal N} \langle e^{\alpha \phi}(i/2)\rangle
\nonumber\\
{\rm residue } \,\,U^R(\alpha) |_{\alpha =  Q/2 - (p+q/2)b}
&= & {\cal N}\langle\sigma  e^{\alpha \phi}(i/2)\rangle,
\eeaq
where $p$ is a non-negative integer and $q$ is a non-negative even
integer for the (NS)-sector and odd for the (R)-sector.
From Eqs.(\ref{UN}) and (\ref{UR}), one can find
nontrivial pole structure of $U^N$ at $p+q/2 = 0 $,
\beq
U^N(\alpha) \cong \frac {\cal N}{\alpha - \frac Q2},
\eeq
and at $p+q/2 = 1$,
\beq
U^N(\alpha) \cong \pi \mu b^2 \gamma ( \frac {bQ} 2)
\Gamma (-b^2 ) \cosh (\pi b s)
\frac {\cal N}{\alpha - \frac Q2+b}.
\eeq

On the other hand, the perturbative calculation for $p+q/2 =0$ gives
\beq
\left< e^{\alpha \phi} (i/2 ,-i /2) \right>_0 =1 \,.
\eeq
and for $p+q/2 =1$
\beaq
\langle e^{\alpha \phi} (i/2,-i/2 ) V_b \rangle_0
&=& \mu b^2 \int_{{\rm Im} z >0  } d^2\xi
|\xi-i/2|^{-2\alpha b} |\xi+i/2|^{- 2\alpha b} |\xi- \bar\xi| ^{-bQ}
\nonumber
\\
&=&
(\pi bQ ) \mu b^2 \Gamma (-bQ) \gamma(\frac{bQ}2)
\\
\langle e^{\alpha \phi} (i/2,-i/2) B_b^2\rangle_0
&=& -\mu_B^2 \int_{-\infty}^{\infty} dx_1 dx_2
|x_1 -i/2|^{-2\alpha b} |x_2 -i/2|^{- 2\alpha b} |x_1- x_2|^{-bQ}
\nonumber
\\
&=&  -\mu_B^2 (4\pi bQ ) \Gamma (-bQ) \gamma\left(\frac{\pi bQ}2\right)
\sin\left(\frac{bQ}2\right).
\eeaq
Combining the two contributions gives the correct residue 
of the one-point function.

The first non-trivial check for the (R)-sector arises at $p+q/2 =1/2$
\beq
U^R (\alpha)\cong \frac {\mu_B \pi }{\alpha - \frac 1{2b} }.
\eeq
This is consistent with the perturbative result
\beq
\left< \sigma e^{\alpha \phi} (i/2 ,-i /2 ) B_b \right>_0
= \mu_B \pi \,.
\eeq

\section{Boundary states for the super-LFT}

For the super-CFTs, Virasoro characters are defined for the 
(NS) sector, the (R)-sector, and the $\widetilde{({\rm NS})}$-sector.
The characters of the primary states for the generic value of $P$,
which have no null-states, are given by \cite{MY}: 
\beaq
\chi^{NS}_P(q^2)&=&\sqrt{{\theta_3(q)\over{\eta(q)}}}{q^{P^2/2}\over{
\eta(q)}}\\
\chi^{\widetilde{NS}}_P(q)&=&
\sqrt{{\theta_4(q)\over{\eta(q)}}}{q^{P^2/2}\over{
\eta(q)}}\\
\chi^{R}_P(q)&=&
\sqrt{{\theta_2(q)\over{2\eta(q)}}}{q^{P^2/2}\over{
\eta(q)}}
\eeaq
where $q=\exp(2\pi i\tau)$.
Under the modular transformation $\tau\to \tau'=-1/\tau$, 
the characters transform
\beaq
\chi^{NS}_P(\tau)&=&\int_{-\infty}^{\infty}dP'e^{-2\pi iPP'}
\chi^{NS}_{P'}(\tau')\\
\chi^{\widetilde{NS}}_P(\tau)&=&\int_{-\infty}^{\infty}dP'e^{-2\pi iPP'}
\chi^{R}_{P'}(\tau')\\
\chi^{R}_P(\tau)&=&\int_{-\infty}^{\infty}dP'e^{-2\pi iPP'}
\chi^{\widetilde{NS}}_{P'}(\tau').
\eeaq

On the other hand, the modular transformations of the characters 
for the (NS) degenerate fields in Eq.(\ref{nsdeg}) are given by
\beaq
\chi^{NS}_{m,n}(q)
&=&\int_{-\infty}^{\infty}dP \chi^{NS}_{P}(q')
2 \sinh (\pi m P/b)  \sinh (\pi m Pb)\label{modsns}  \\
\chi^{\widetilde{NS}}_{m,n}(q)
&=&\int_{-\infty}^{\infty}dP \chi^{R}_{P}(q')
\left\{ \begin{array}{l}
2 \sinh (\pi m P/b)\sinh (\pi m Pb) \qquad  m,n = {\rm even} \\
2 \cosh (\pi m P/b)\cosh (\pi m Pb)  \qquad  m,n = {\rm odd.} 
\end{array} 
\right.
\label{modsr}
\eeaq

According to Cardy's formalism, one can associate
a conformal BC with each primary state \cite{Cardy}. 
Since the SLFT is an irrational CFT with infinite number of primary
states, there will be infinite number of conformal BCs.
One can classify these into `discrete BCs' and `continuous BCs'
for the degenerate and non-degenerate primary states, respectively.
It is natural to start with the discrete BCs $(m,n)$.
With $m-n=$even, there are the (NS)-type BCs corresponding to the
(NS) degenerate fields $N_{\alpha_{m,n}}$.
We will denote the corresponding boundary states by $|(m,n)\rangle$.
Due to the super conformal symmetry, one needs to introduce
additional BCs $(\widetilde{m,n})$ and corresponding boundary states
$|(\widetilde{m,n})\rangle$ \cite{Nepomechie}. 
Let us consider $(m,n)$ BCs first.
Through the modular transformation, one can obtain
\beq
\chi^{NS}_{m,n}(\tau) =\int_{-\infty}^{\infty}dP
\Psi_{m,n}^{NS}(P) \Psi_{1,1}^{NS}(P)^{\dagger}
\chi^{NS}_{P}(\tau'),\label{nscardy}
\eeq
where the amplitude is defined by
\[
\Psi_{m,n}^{NS}(P)=\langle (m,n)|\alpha,{\rm NS}\rangle\rangle,
\]
where $|\alpha,NS\rangle\rangle$ is the (NS) Ishibashi state with
$\alpha=Q/2+iP$.

Using Eq.(\ref{modsns}) into Eq.(\ref{nscardy}), one can obtain 
a relation
\beq
\Psi_{1,1}^{NS}(P)^{\dagger}\Psi_{m,n}^{NS}(P)
=2\sinh(\pi m P/b)\sinh(\pi m Pb).
\eeq
We can set the basic amplitude for $m=n=1$ as
\beq
\Psi^{NS}_{1,1}(P) = {\pi \sqrt{2}
\over P\Gamma(-iPb)\Gamma(-iP/b) }
\left[\pi\mu\gamma\left({bQ\over{2}}\right)\right]^{-iP/b}.
\eeq
Then, one can obtain the relation
\beq
\Psi^{NS}_{m,n} (P) =\Psi^{NS}_{1,1} (P) \,\,
{\sinh\left({m\pi P\over{b}}\right)
\sinh\left({n\pi P b}\right)
\over{
\sinh\left({\pi P\over{b}}\right)
\sinh\left({\pi P b}\right)}}
\label{ratio}
\eeq
by the Cardy formula.

It has been shown in \cite{CarLew} that the amplitudes are related 
to the one-point functions by
\beq
U_{\tilde{k}}(\phi)={\langle\tilde{k}|\phi\rangle\rangle\over{
\langle\tilde{k}|{\bf 1}\rangle\rangle}}.
\label{carlew}
\eeq
For the SLFT, this formula becomes
\beq
U^{N}_{m,n}(\alpha)={\langle (m,n)|\alpha,{\rm NS}\rangle\rangle\over{
\langle (m,n)|0,{\rm NS}\rangle\rangle}}.
\eeq
From this and Eq.(\ref{ratio}), one can obtain
\beq
{U^{N}_{m,n}(\alpha)\over{U^{N}_{1,1}(\alpha)}}=
{\langle (m,n)|\alpha,{\rm NS}\rangle\rangle\over{
\langle (m,n)|0,{\rm NS}\rangle\rangle}}
{\langle (1,1)|0,{\rm NS}\rangle\rangle\over{
\langle (1,1)|\alpha,{\rm NS}\rangle\rangle}}
={\sinh\left({m\pi P\over{b}}\right)
\sinh\left({n\pi P b}\right)
\over{
\sinh\left({\pi P\over{b}}\right)
\sinh\left({\pi P b}\right)}}
{\sin\left({\pi Q\over{2b}}\right)
\sin\left({\pi Qb\over{2}}\right)
\over{
\sin\left({m\pi Q\over{2b}}\right)
\sin\left({n\pi Qb\over{2}}\right)}}.
\eeq
This result is identical to Eq.(\ref{unmn}), which shows that 
the one-point functions obtained from the functional relations
are consistent with the modular transformation properties.

Now let us consider the partition function on a strip
with $(m,n)$ and $(m',n')$ BCs on both boundaries.
Using the fusion procedure, one can obtain
\beaq
Z^{NS}_{(m,n), (m' ,n')}(\tau)
&=&\int_{-\infty}^{\infty} dP
\chi^{NS}_{P}(\tau' )
\Psi^{NS}_{m,n} (P)  \Psi^{NS}_{m',n'} (P)^{\dagger} \,.
\nonumber\\
& =&\int_{-\infty}^{\infty} dP
\chi^{NS}_{P}(\tau' )
\frac{
2 \sinh(m\pi P/b) \sinh(n\pi Pb)
\sinh(m'\pi P/b) \sinh(n'\pi Pb)}
{\sinh(\pi P/b) \sinh(\pi Pb)  }
\nonumber\\
&=& \sum_{k=0}^{{\rm min}(m,m')-1}
\sum_{l=0}^{{\rm min}(n,n')-1}
\chi_{(m+m'-1-2k),(n+n'-1-2l)}^{NS}(\tau)
\eeaq
in accordance with the fusion algebra.
The character for the non-degenerate case with $P=s/2$ 
satisfies
\beq
\chi^{NS}_{s}(\tau)
=\int_{-\infty}^{\infty}dP
\chi^{NS}_{P}(\tau')
\cos (\pi s P )
=\int_{-\infty}^{\infty}dP
\chi^{NS}_{P}(\tau')
\Psi_{s}^{NS}(P) \Psi_{1,1}^{NS}(P)^{\dagger},
\eeq
with $\Psi_{s}^{NS}(P)=\langle\tilde{s}|P,NS\rangle\rangle$.

From this, one can find the amplitudes for a general non-degenerate (NS) 
boundary state satisfy
\beq
\Psi^{NS}_{s} (P) =\Psi^{NS}_{1,1} (P) \,\,
{\cos( \pi s P) \over
2\sinh\left({\pi P\over{b}}\right)
\sinh\left({\pi P b}\right)}\,.
\eeq
Again, from Eq.(\ref{carlew}), 
the amplitudes are related to the one-point functions by
\beq
{U^{N}(\alpha)\over{U^{N}_{1,1}(\alpha)}}=
{\langle\tilde{s}|\alpha,{\rm NS}\rangle\rangle\over{
\langle\tilde{s}|0,{\rm NS}\rangle\rangle}}
{\langle (1,1)|0,{\rm NS}\rangle\rangle\over{
\langle (1,1)|\alpha,{\rm NS}\rangle\rangle}}
={\cos(\pi Ps)\over{\cosh\left({Q\pi s\over{2}}\right)}}
{\sin(\pi Q/2b) \sin(\pi Qb/2)\over{\sinh(\pi Pb)\sinh(\pi P/b)}}.
\eeq
This can be checked to be correct by Eqs.(\ref{un11}) and (\ref{UN}).

The partition function with a discrete BC on one side and a continuous
BC on the other is given as follows:
\beaq
Z^{NS}_{(m,n), s}(\tau)
&=&\int_{-\infty}^{\infty} dP
\chi^{NS}_{P}(\tau' )
\Psi^{NS}_{m,n} (P)  \Psi^{NS}_{s}(P)^{\dagger} 
\nonumber\\
&=& \int_{-\infty}^{\infty} dP
\chi^{NS}_{P}(\tau' )
{
\sinh(m\pi P/b) \sinh(n\pi Pb)
\over
\sinh(\pi P/b) \sinh(\pi Pb)  
} \cos( \pi s P) 
\nonumber\\
&=& \sum_{k=0}^{m-1}
\sum_{l=0}^{n-1}
\chi_{s +i (m-1-2k)/b +i (n-1-2l)b}^{NS}(\tau)\,,
\eeaq
which goes with the fusion algebra.

The partition function $Z^{NS}_{s, s'}(\tau)$
with contiuous BCs on both boundaries, $s$ and $s'$, 
is given as
\beq
Z^{NS}_{s, s'}(\tau)
=\int_{-\infty}^{\infty} dP
\chi^{NS}_{P}(\tau' )
\Psi^{NS}_{s} (P)  \Psi^{NS}_{s'} (P)^{\dagger}\,.
\eeq
This can be rewritten as 
\beaq
Z^{NS}_{s, s'}(\tau)
&=& \int_{-\infty}^{\infty} dP'
\int_{-\infty}^{\infty} dP
e^{- 2 i \pi P P'}
\chi^{NS}_{P }(\tau )
\Psi^{NS}_{s} (P')  \Psi^{NS}_{s'} (P')^{\dagger} \,.
\nonumber\\
&=& \int_{0}^{\infty} dP
\chi^{NS}_{P }(\tau )
\rho^{NS}_{ss'}(P),
\eeaq
where $\rho^{NS}_{ss'}(P)$ is the density of states,
\beq
\rho^{NS}_{ss'}(P)=\int_{-\infty}^{\infty} { dt \over \pi }
e^{- 2 i P t}
{\cos (s t) \cos ( s' t) 
\over
\sinh (t/b ) \sinh (t b) }.
\eeq
This quantity is not well-defined at $P=0$ 
and is to be properly regularized.  
This density of states is, on the other hand, conjectured to be 
related with the boundary two-point function of
$n_{\beta}^{ss'}$ with $\beta=Q/2+iP$, $d_B^{NS}(P|s,s')$, by
\beq
\rho^{NS}_{s,s'} (P) = - {i \over 2\pi } 
{d \over dP} \log d^{NS}_B (P|s,s').
\eeq
It is remarkable that this relation is identical to that of the LFT.
This means that the boundary (NS) two-point functions have the
same dependence on the boundary parameters as the LFT which is 
obtained in \cite{FZZ}.

Now we consider the (R) operator propagating in the strip.
From the relation
\beq
\chi^{\widetilde{NS}}_{m,n}(\tau)
=\int_{-\infty}^{\infty}dP
\Psi_{m,n}^{R}(P) \Psi_{1,1}^{R}(P)^{\dagger}
\chi^{R}_{P}(\tau'),
\eeq
one can obtain
\beq
\Psi_{1,1}^{R}(P)^{\dagger}\Psi_{m,n}^{R}(P)=
\left\{ \begin{array}{l}
2 \sinh (\pi m P/b)\sinh (\pi m Pb) \qquad  m,n = {\rm even} \\
2 \cosh (\pi m P/b)\cosh (\pi m Pb)  \qquad  m,n = {\rm odd.}
\end{array}
\right.
\label{ramp}
\eeq
Here, we define the amplitude
$\Psi_{m,n}^{R}(P)=\langle (m,n)|\alpha,R\rangle\rangle$ as before.

These amplitudes are related to the one-point fuctions by
\beq
U^{R}_{m,n}(\alpha)={\langle (m,n)|\alpha,R\rangle\rangle\over{
\langle (m,n)|0,R\rangle\rangle}},
\eeq
and satisfy
\beq
{U^{R}_{m,n}(\alpha)\over{U^{R}_{1,1}(\alpha)}}=
{\langle (m,n)|\alpha,R\rangle\rangle\over{
\langle (m,n)|0,R\rangle\rangle}}
{\langle (1,1)|0,R\rangle\rangle\over{
\langle (1,1)|\alpha,R\rangle\rangle}}.
\eeq
One can check that this relation is consistent with Eqs.(\ref{ramp}) and 
(\ref{urmn}). 
This shows again that
the functional relations are consistent with the modular transformation.

The partition function with $(m,n)$ and $(m',n')$ BCs is written as
\beaq
\nonumber
&&Z^{\widetilde{NS}}_{(m,n), (m' ,n' )}(\tau)
=\int_{-\infty}^{\infty} dP
\sqrt2 \chi^{R}_{P}(\tau' )\,.
\Psi^{R}_{m,n}(P)  \Psi^{R}_{m',n'}(P)^{\dagger}\\
&& \quad
= \sum_{k=0}^{{\rm min}(m,m')-1}
\sum_{l=0}^{{\rm min}(n,n')-1}
\chi_{(m+m'-1-2k),(n+n'-1-2l)}^{\widetilde{NS}}(\tau)
\eeaq
and is consistent with fusion algebra.

The character for the continuous boundary parameter $s$ satisfies
\beaq
\chi^{\widetilde{NS}}_{s}(\tau)
&=&\int_{-\infty}^{\infty}dP
\chi^{R}_{P}(\tau')
\cos (\pi s P)
=\int_{-\infty}^{\infty}dP
\chi^{R}_{P}(\tau')
\Psi_{s}^{R}(P)\Psi_{1,1}^{R}(P)^{\dagger}.
\eeaq
The (R) amplitude for a general boundary parameter $s$ is given as
\beq
\Psi^{R}_{s} (P) =\Psi^{R}_{1,1} (P) \,\,
{\cos( \pi s P) \over
2\cosh\left({\pi P\over{b}}\right)
\cosh\left({\pi P b}\right)}.
\eeq
Again, from Eq.(\ref{carlew}),
the amplitudes are related to the one-point functions by
\beq
{U^{R}(\alpha)\over{U^{R}_{1,1}(\alpha)}}=
{\langle\tilde{s}|\alpha,R\rangle\rangle\over{
\langle\tilde{s}|0,R\rangle\rangle}}
{\langle (1,1)|0,R\rangle\rangle\over{
\langle (1,1)|\alpha,R\rangle\rangle}}
={\cos(\pi Ps)\over{\cosh\left({Q\pi s\over{2}}\right)}}
{\cos(\pi Q/2b) \cos(\pi Qb/2)\over{\cosh(\pi Pb)\cosh(\pi P/b)}}.
\eeq
This can be checked to be correct by Eqs.(\ref{ur11}) and (\ref{UR}).

The partition function with mixed BCs is given as follows:
\beaq
Z^{\widetilde{NS}}_{(m,n), s}(\tau)
&=&\int_{-\infty}^{\infty} dP
\chi^{R}_{P}(\tau' )
\Psi^{R}_{m,n} (P)  \Psi^{R}_{s}(P)^{\dagger} 
\nonumber\\
&=& \sum_{k=0}^{m-1}
\sum_{l=0}^{n-1}
\chi_{s +i (m-1-2k)/b +i (n-1-2l)b}^{\widetilde{NS}}(\tau)\,.
\eeaq

One can consider $(\widetilde{m,n})$ BCs in the same way and
can associate the amplitudes $\Psi^{NS}_{\widetilde{m,n}}$ and 
$\Psi^{R}_{\widetilde{m,n}}$
with ${\tilde U}^{N}_{m,n}$ and ${\tilde U}^{R}_{m,n}$, respectively.
From Eq.(\ref{utilde}), one can see that
\beq
\Psi^{NS}_{m,n}=\Psi^{NS}_{\widetilde{m,n}},\quad
\Psi^{R}_{m,n}=-\Psi^{R}_{\widetilde{m,n}}.
\eeq
Similar result holds for the continuous BCs and can be compared with
Eq.(\ref{utildec}).
These results are consistent with those of the rational super-CFTs 
considered in \cite{Nepomechie}.

\section{Conclusions}
In this paper, we have studied the SLFT in two-dimensional space-time
with boundary applying the same method used for the LFT.
However, the one-point functions of the SLFT 
satisfy more complicated functional relations
due to the existence of two sectors, the (NS) and (R).
By solving the functional relations, we find not only the one-point functions
but also the relation between the parameter $\mu_B$ in the boundary
action and that for the boundary condition.

We have also related the one-point functions in the pseudosphere to
the conformal BCs and showed that they are consistent with the Cardy
formalism if one takes care of the peculiar aspects of the super-CFTs
in the same way as the rational cases \cite{Apikyan,Nepomechie}.
Then, this result has been used to understand boundary two-point functions
for the SLFT with the boundary action.
We conclude that the boundary (NS) two-point functions have the 
same dependence on the boundary parameters as the LFT while
explicit expression of this quantity needs more work.
We hope to present it in another publication.

There are still other problems which should be further explored.
The solutions of Eqs.(\ref{unmn}) and (\ref{urmn}) are possible only for
$m-n$=even. 
The functional equations for the other case may also exist.
This becomes necessary when one relates to the conformal BCs.
In this paper, we considered only the (NS)-type BCs since the (R)-type
BCs should be associated with $m-n$=odd.

Based on the successful results of the LFT and the SLFT,
it seems the approach based on the functional relation associated with
some degenerate fields are quite efficient way of dealing with irrational
CFTs. 
In this respect, it would be interesting to apply this method to such
irrational CFTs as finite Toda field theories and the LFT with 
$N=2$ and fractional supersymmetries.
Another interesting problem is to derive the boundary reflection 
amplitude from the boundary two-point functions 
of the $N=1$ SLFT and to obtain off-critical scaling function developed in
\cite{AKR}.

\section*{\bf Acknowledgement}

We thank V. Fateev, R. Nepomechie and Al. Zamolodchikov for valuable 
discussions. 
We thank Univ. Montpellier II, APCTP and YVRC for hospitality. 
This work is supported in part by KOSEF 1999-2-112-001-5, 
MOST-99-N6-01-01-A-5 (CA), and Eastern Europe exchange program 12-69-002 
sponsored by KISTEP (MS).


\begin{thebibliography}{99}
\bibitem{Gervais} G.-L. Gervais, Comm. Math. Phys. {\bf 130} (1990) 252.
\bibitem{Teschner} J. Teschner, Phys. Lett. {\bf B363} (1995) 65.
\bibitem{ZamZam} A. B. Zamolodchikov and Al. B. Zamolodchikov,
Nucl. Phys. {\bf B477} (1996) 577.
\bibitem{RasSta} R. C. Rashkov and M. Stanishkov,
Phys. Lett. {\bf B380} (1996) 49.
\bibitem{Poghossian} R. H. Poghossian, Nucl.Phys. {\bf B496} (1997) 451.
\bibitem{Cardy} J. Cardy, Nucl. Phys. {\bf B240} (1984) 514.
\bibitem{string} P. Lee, H. Ooguri, and J. Park, ``Boundary States for
${\rm AdS}_2$ branes in ${\rm AdS}_3$'', {\tt hep-th/0112188};
B. Ponsot, V. Schomerus, and J. Teschner, ``Branes in the Euclidean
${\rm AdS}_3$'', {\tt hep-th/0112198}.
\bibitem{FZZ} V. A. Fateev, A. B. Zamolodchikov, and 
Al. B. Zamolodchikov, ``Boundary Liouville Field Theory I.
Boundary State and Boundary Two-point Function'', {\tt hep-th/0001012}.
\bibitem{ZamZam2} A. B. Zamolodchikov and Al. B. Zamolodchikov,
``Liouville Field Theory on a Pseudosphere'', {\tt hep-th/0101152}.
\bibitem{CarLew} J. Cardy and D. Lewellen, Phys. Lett. {\bf B259} 
(1991) 274.
\bibitem{Apikyan} S.A. Apikyan and D.A. Sahakyan, Mod. Phys. Lett.
{\bf A14} (1999) 211.
\bibitem{Nepomechie} R. Nepomechie, J. Phys. {\bf A34} (2001) 6509.
\bibitem{FQS} D. Friedan, Z. Qiu, S. Shenker,
Phys. Lett. {\bf B151} (1985) 37.
\bibitem{MSS} G. Musssardo, G. Sotkov, and M. Stanishkov, Nucl.Phys.
{\bf B305} (1988) 69. 
\bibitem{GZ} S. Ghoshal, A. Zamolodchikov, Int. J. Mod. Phys. 
{\bf A9} (1993) 3841.
\bibitem{Prata} J. Prata, Phys. Lett. {\bf B405} (1997) 271.
\bibitem{MY} Y. Matsuo and S. Yahikozawa, Phys. Lett. {\bf B178} (1986) 211; 
D. Kastor, Nucl. Phys. {\bf B280} (1987) 304.
\bibitem{AKR} C. Ahn, C. Kim, and C. Rim, Nucl. Phys. {\bf B556} (1999) 505;
``Reflection Amplitudes of Boundary Toda Theories and Thermodynamic
Bethe ansatz'', {\tt hep-th/0110218}.

\end{thebibliography}
\end{document}